\documentclass[transmag,journal]{IEEEtran}

\usepackage[compress]{cite}
\usepackage{graphicx}
\usepackage{amssymb,amsmath,upgreek,url,lipsum,bm}
\usepackage[per-mode=symbol]{siunitx}
\usepackage{svg}
\usepackage{adjustbox}
\usepackage{flushend}
\usepackage{xcolor}
\usepackage{bm}
\usepackage{svg}
\usepackage{comment}
\usepackage{siunitx}
\usepackage{pdfpages}
\usepackage{booktabs}
\usepackage{makecell}
\usepackage{yhmath}
\usepackage{float}

\newcommand{\sccm}{$\mathrm{sccm}$}

\newcommand{\mr}{\mathrm}
\DeclareSIUnit{\bar}{bar}
\DeclareSIUnit\angstrom{\text{\r{A}}}
\DeclareSIUnit{\dBm}{\text{dBm}}

\begin{document}
%

\title{Design and Realization of Broadband Magnonic Spectrometers \\With Local Electrical Outputs}


\author{
    \IEEEauthorblockN{Johannes Greil\IEEEauthorrefmark{1}, Maximilian Hofschen\IEEEauthorrefmark{1}, Felix Naunheimer\IEEEauthorrefmark{1}, \'{A}d\'{a}m Papp\IEEEauthorrefmark{2}, György Csaba\IEEEauthorrefmark{2}, Markus Becherer\IEEEauthorrefmark{1}
    }
    \IEEEauthorblockA{\IEEEauthorrefmark{1}School of Computation, Information and Technology, Technical University of Munich, München, Germany}
    \IEEEauthorblockA{\IEEEauthorrefmark{2}Faculty of Information Technology and Bionics, P\'{a}zm\'{a}ny P\'{e}ter Catholic University, Budapest, Hungary}
}

\IEEEtitleabstractindextext{%
\begin{abstract}
Microscopic radio-frequency (RF) devices based on propagating spin waves (SWs) are promising for compact, energy-efficient RF signal processing, but their implementation is impeded by fabrication complexity and the lack of efficient electrical readout. 
In this work, we demonstrate a SW-based Rowland circle spectrometer with electrical input and local electrical output transducers. 
The device is realized using a scalable fabrication process based on sputter deposition and wet-chemical etching of Yttrium–Iron–Garnet (YIG), forming concave grating structures with micrometer-scale features.
The device functionality is confirmed by combined electrical and magneto-optical measurements, which show that the deflection of SW wavefronts at different input frequencies closely follows the analytically predicted behavior.
The linear excitation of SWs via two input tones further confirms the spectrometer operation for simultaneously propagating waves.
Beyond the single-device demonstration, we propose a concept for scalable architectures comprising multiple Rowland circles with tunable operating points. 
When combined with broadband parallel electrical readout, this approach enables control over bandwidth and spectral resolution, which are relevant to spectral occupancy detection in wireless communication systems.
\end{abstract}
}

\maketitle
\section{Introduction}

In the context of emerging 5G and future 6G communication systems, the demand for compact and efficient microwave signal processing devices continues to increase~\cite{Levchenko2026}.
Magnonic devices using spin waves (SWs), with their quasi-particles called magnons, as information carriers are promising candidates due to their potential for low power consumption, high operational frequencies, and broad tunability ranges~\cite{Flebus2024,Wang2024}.
Particularly relevant are spatially dispersive~\cite{Davies2015,Whitehead2019,Vogel2020,Kiechle2022,Kiechle2023} and wavevector-selective devices~\cite{Yagan2021,Taniguchi2022}, which enable functionalities such as spectrum analysis.

In this context, the Rowland circle, originally developed for optical spectrometers, has been proposed in~\cite{Papp2017} as a concept for achieving wavefront separation in magnonic systems.
Previous work has demonstrated Rowland circle-based magnonic spectrometers through simulations~\cite{Papp2017} and an experimental implementation~\cite{Papp2021}.
In the latter, a continuous Yttrium-Iron-Garnet (YIG) film was employed together with a straight excitation transducer and gratings defined by focused-ion-beam (FIB) irradiation, while device characterization relied on magneto-optical measurements to verify the functionality.

In the present work, we extend this approach to a realization compatible with electrical interfacing.
A structured YIG island, obtained by wet-chemical etching, is used to reduce undesired parasitic magnonic signal contributions from the feed lines. 
Furthermore, a curvilinear excitation transducer enables localized, coherent SW emission across the grating, thereby supporting the linear excitation condition required for spectrometer operation. 
Most importantly, the spatially separated SW wavefronts are accessed via integrated local electrical output transducers, enabling all-electrical detection of spatially resolved magnonic signals. 
The functionality of the spectrometer is not only confirmed by single-tone excitation, but also demonstrated using two-tone measurements, which is a prerequisite for practical spectrometer operation.

Building on the experimental demonstration, we outline a concept for scalable spectrometer architectures based on parallelized Rowland circles with adjustable operational points. 
Furthermore, the integration of a broadband, parallel electrical readout scheme is suggested.
Implications for RF front-end signal processing are briefly discussed, including potential applications in listen-before-talk (LBT) protocols~\cite{Zhang2015,Lagen2020,Ma2020,Baswade2021,Brauer2021} and wake-up receiver architectures\cite{Hoglund2024}.

\section{Rowland Circle Concept and Device Design}

\begin{figure}[h!]
    \includegraphics[width=\columnwidth]{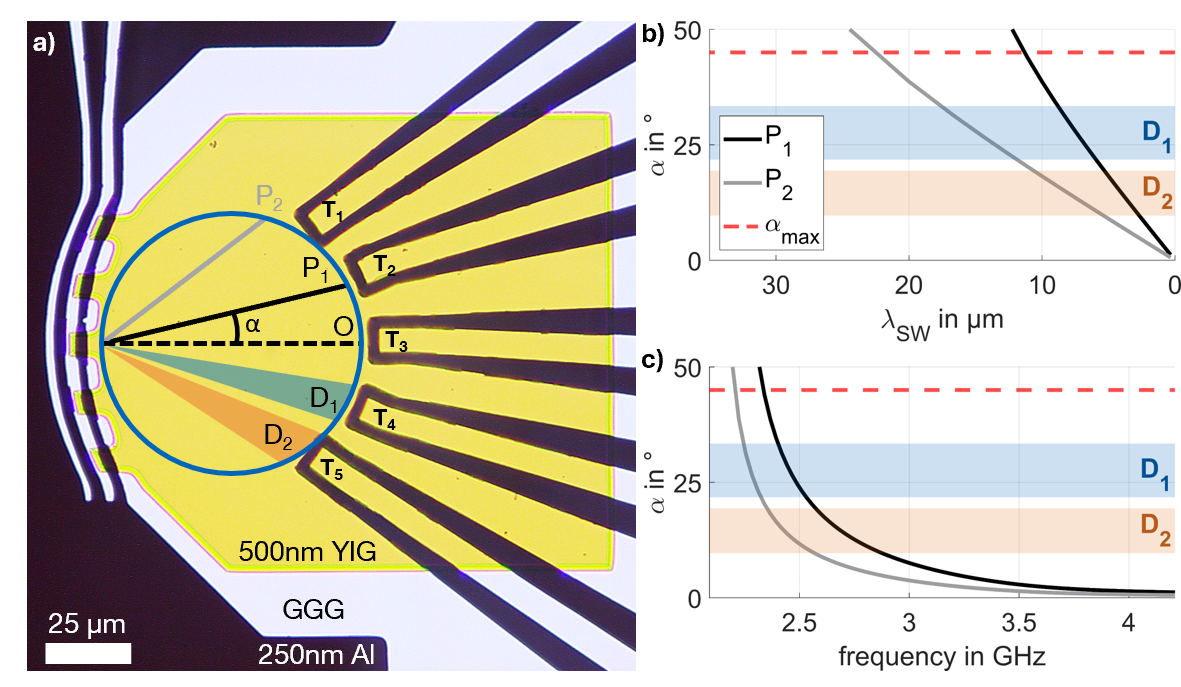}
    \caption{\textbf{Realization of a magnonic Rowland circle with local electrical read-out.} a)~Through-light microscope image of a fabricated Rowland circle spectrometer with radius $R=\SI{75}{\micro\meter}$ and grating period $d=\SI{16}{\micro\meter}$. A concave YIG grating is realized by wet-chemical etching, and SWs are excited by a curvilinear transducer. SW interference leads to spatial separation of different diffraction orders along the Rowland circle, with the zero-order focused at $O$ and higher orders deflected to $P_n$ according to $\alpha = \arcsin(n\lambda/d)$. Local microstrip transducers ($T_1$--$T_5$) enable electrical detection. b)~Calculated deflection angle $\alpha$ as a function of wavelength $\lambda$ for diffraction orders $n=1,2$, including the practical limit at $\alpha=\SI{45}{\degree}$ (red dashed line). c)~Corresponding dependence of $\alpha$ on excitation frequency $f$, exhibiting non-linearity (with frequency) due to SW dispersion. While equal angular intervals correspond to uniform wavelength spans, they result in non-uniform frequency coverage, affecting the bandwidth of the detection windows $D_1$ and $D_2$.}
    \label{fig:device_concept}
\end{figure}

The device concept is based on the Rowland circle geometry, which enables wavefront separation via wavelength-dependent diffraction in a concave grating~\cite{James2007}.
The implementation follows the principle introduced in~\cite{Papp2017}, where a structured YIG island, a linear excitation transducer, and spatially distributed output transducers form a magnonic spectrometer.
For a grating with period $d$ and large grating radius $R$, the deflection angle $\alpha$ of a wavefront with wavelength $\lambda$ is given by~\cite{Papp2017}
\begin{equation}
    \label{eq:DeflectionAngle}
    \alpha = \arcsin\left(\frac{n\lambda}{d}\right),
\end{equation}
with diffraction order $n$ (see Fig.\ref{fig:device_concept}a). 
Wavefronts are focused along the Rowland circle with radius $r=R/2$, where the zero-order mode is detected at the reference position $O$, while first- and higher-order modes are spatially separated at positions $P_n$. 

Figure~\ref{fig:device_concept}b shows the dependence of $\alpha$ with $\lambda$ for the first and second diffraction orders ($n=1,2$), including the practical focusing limit at $\alpha=\SI{45}{\degree}$ (red dashed line). 
In contrast, Fig.~\ref{fig:device_concept}c) depicts the variation of $\alpha$ with frequency $f$, where the SW dispersion leads to a pronounced non-linear change with $f$ (see Appendix~\ref{sec:A_ModelsAndCalcs}).
This difference is relevant to device operation, as the design is performed in the wavelength domain, whereas the signal processing task is ultimately performed in the frequency domain. 
Detection windows that are defined by equal intervals in $\alpha(\lambda)$ correspond to unequal frequency intervals.
This results in unequal frequency coverage for identical geometrical detection windows $D_1$ and $D_2$, which are defined by pairs of symmetric transducers ($T_1$/$T_5$ and $T_2$/$T_4$).
The determination of the detection windows is based on the analytical Kalinikos-Slavin SW dispersion relation~\cite{Kalinikos1986} and the transducers' excitation function $\rho$~(see Appendix~\ref{sec:A_ModelsAndCalcs}).

Compared to the realization in~\cite{Papp2021}, the present implementation introduces several modifications to enable electrical interfacing and linear SW excitation.
The concave grating is realized by wet-chemical etching of a sputtered YIG film, forming a patterned island with YIG areas added to all non-grating sides to prevent reflections. 
Excitation of SWs is achieved using a curvilinear transducer integrated on top of the grating. 
In contrast to the $\lambda/2$ modulation depth proposed in~\cite{Papp2017}, the excitation is localized to the grating ridges, resulting in an array of closely spaced, quasi-point-like and coherent SW sources, as the electromagnetic wavelength is orders of magnitude larger than the transducer length.
This concept mimics the local blocking of incident waves to the grating as it would be realized in optics.
The loop-shaped output transducers proposed in~\cite{Papp2017} are simplified to short straight microstrip line segments, thereby defining localized detection positions along the Rowland circle. 
The output transducers are connected to one ground-signal (GS) pair of coplanar waveguides (CPWs, in a ground-signal-ground (GSG) configuration) and operate as shorted terminations, enabling all-electrical detection of the spatially separated wavefronts.


\section{Methods}
\subsection{Sample Fabrication}

\begin{figure}[t!]
    \centering
    \includegraphics[width=\columnwidth]{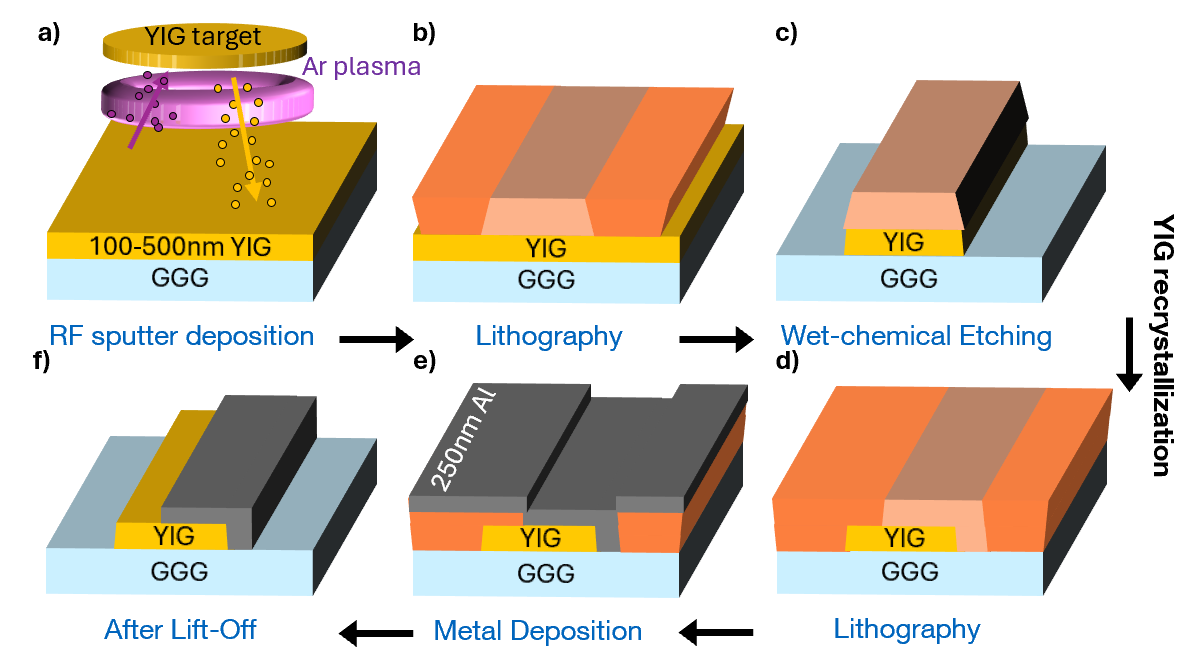}
    \caption{\textbf{Sample Fabrication using sputtering and wet-chemical etching.} a)~RF magnetron sputtering of an amorphous YIG film (\SIrange{100}{500}{\nano\meter}) in an Ar plasma. b)~Optical lithography using a maskless aligner to define an etching mask. c)~Wet-chemical etching of the amorphous YIG film using a standard solution of phosphoric, nitric, and acetic acid at \SI{45}{\celsius} with an etch rate of \SI{0.87}{\nano\meter\per\s}. d)~Optical lithography using a maskless aligner to define a lift-off mask. e)~Thermal evaporation of \SIrange{200}{250}{\nano\meter} Aluminum. f)~After lift-off, the YIG islands, including the grating structures, and the GGG substrate are partially metalized.}
    \label{fig:fabrication}
\end{figure}

\begin{figure*}[h!]
    \centering
    \includegraphics[width=\linewidth]{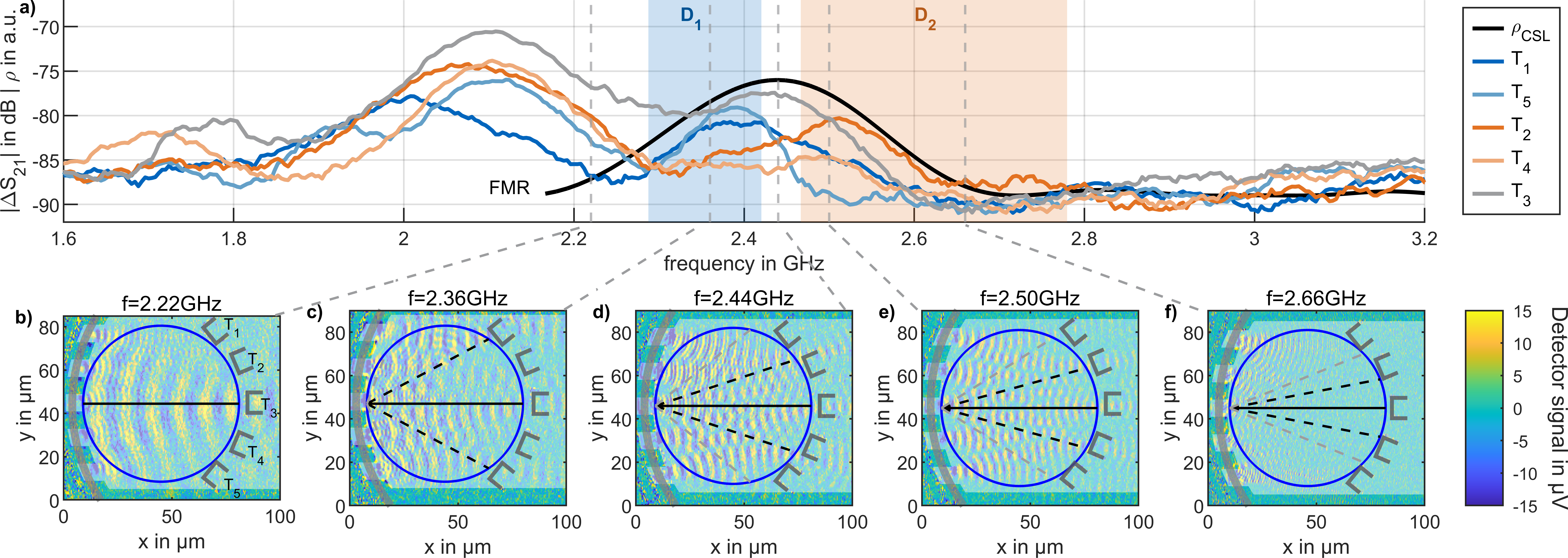}
    \caption{\textbf{Electrical and magneto-optical characterization of a magnonic Rowland circle} a)~Calibrated  differential transmission S-parameters $|\Delta S_{21}|$ for a concave grating with $R=\SI{75}{\micro\meter}$ and $d=\SI{16}{\micro\meter}$ at a constant bias field $\mu_0H_x \approx \SI{246}{\milli\tesla}$. The measured detection peaks for the symmetrical output transducers $T_1/T_2$ and $T_4/T_5$ are located in the analytically calculated detection regions $D_{1,2}$. The zero-order focusing point $O$ serves as the reference signal at~$T_3$. The excitation function for the used input transducer geometry can be approximated by a coplanar strip (CSL) line and is overlaid on the measurements. b)-f)~trMOKE measurements at the same bias field. The concave grating, input/output transducers, and theoretical deflection angles for the first- and second-order diffraction are overlaid on the colormaps. While in b) only the zero-order component is detected, in c) detection span $D_1$ is focused. For shorter wavelengths (higher frequencies), there is a gap in the detection regions (\SI{2.44}{\giga\hertz}) in d), followed by the detection in span $D_2$ in e), where the second-order wavefront is deflected to $T_1/T_5$, marking the upper limit of the unambiguous frequency detection range. f)~Shorter wavelengths are less efficiently excited, and the damping per wavelength limits the maximum propagation distance.}
    \label{fig:single_tone}
\end{figure*}

A schematic overview of the fabrication process is shown in Fig.~\ref{fig:fabrication}. 
YIG films with thicknesses of \SI{100}{\nano\meter} or \SI{500}{\nano\meter} are deposited by RF magnetron sputtering on Gadolinium–Gallium–Garnet (GGG) substrates, which are cleaned both wet-chemically and by Ar plasma sputtering at \SI{50}{\watt} for \SI{3}{\minute}. 
The deposition is carried out in an Ar plasma at a base pressure of \SI{3e-7}{\milli\bar}, a working pressure of \SI{8}{\micro\bar}, an Ar flow of \SI{25}{}\,\sccm, and an RF power of \SI{100}{\watt}, yielding a growth rate of \SI{8.5}{\nano\meter\per\minute}. 
The deposited films are initially amorphous.
Patterning is performed by optical lithography using a maskless aligner to define a resist mask, followed by wet-chemical etching in a solution of $H_3PO_4:HNO_3:CH_3COOH:H_2O=(80:5:5:10)\%$ (TechniEtch Al80) at \SI{45}{\celsius} with an etch rate of \SI{0.87}{\nano\meter\per\s}. 
The isotropic nature of the process results in a lateral resolution loss on the order of the film thickness, but remains negligible for the micrometer-scale structures in this work.
After resist removal, the structured YIG islands are recrystallized at \SI{700}{\celsius} for \SI{4}{\hour} in an oxidation furnace under a slight oxygen flow of \SI{0.5}{\liter\per\minute}. 
Heating is performed at \SI{10}{\celsius\per\minute}, while controlled cooling is critical for the film quality and carried out at \SI{2}{\celsius\per\minute} down to \SI{400}{\celsius} and subsequently at \SI{1}{\celsius\per\minute} to room temperature. 
Broadband flip-chip ferromagnetic resonance measurements confirm that the etching process does not alter the magnetic properties compared to sputtered continuous films, yielding typical values of $M_\mathrm{eff}=\SI{133}{\kilo\ampere\per\meter}$ and a Gilbert damping of $\alpha_\mr{G}=\SI{4e-4}{}$.
In a second lithography step, a lift-off mask is defined for metallization, followed by thermal evaporation of \SIrange{200}{250}{\nano\meter} thick Aluminum at \SI{5.5e-7}{\milli\bar} and a deposition rate of approximately \SI{4}{\angstrom\per\second}.

\subsection{Measurement Setups}

Optical characterization was performed using a time-resolved magneto-optical Kerr-effect (trMOKE) microscope based on sub-Nyquist sampling (periodic undersampling). 
Measurements in the forward-volume (FV) geometry were enabled by a height-adjustable permanent magnet. 
The focused laser spot has a diameter of approx. \SI{1.5}{\micro\meter}. 
Combined with a lateral scan step size of about \SI{100}{\nano\meter}, this allows reliable detection of spin waves with wavelengths down to \SI{1}{\micro\meter}.

RF characterization was carried out in a wafer prober equipped with an integrated electromagnet using a vector network analyzer (VNA). 
Ground-signal-ground (GSG) probes with a pitch of \SI{150}{\micro\meter} were used to contact the CPW landing pads. 
The VNA reference plane was calibrated to the probe tips using a Through-Open-Short-Match (TOSM) procedure. 
Measurements were performed in the range from \SIrange{1}{6}{\giga\hertz} with a frequency step size of $\Delta f=\SI{1}{\mega\hertz}$, an intermediate-frequency bandwidth (IFBW) of \SI{300}{\hertz},without averaging or smoothing, and for input powers from \SIrange{-15}{10}{\dBm}. 
Additional purely electrical measurements were conducted up to \SI{10}{\giga\hertz}. 
All measurements were performed using a single pair of GSG probes. Multi-port responses were obtained by sequentially reconnecting the output probe (port~2) to the respective local output transducers.


\section{Device Characterization}
\label{sec:Characterization}

\subsection{Single-Tone Excitation}
\label{ssec:Characterization_SingleTone}
For a direct comparison of trMOKE and VNA measurements, a center frequency of \SI{2.45}{\giga\hertz} in an ISM band~\cite{ITU} is considered. 
Figure~\ref{fig:single_tone}a shows the differential transmission $\Delta S_{21}=S_{21,H_x}-S_{21,H_\mathrm{ref}}$ with $\mu_0H_x\approx\SI{246}{\milli\tesla}$ and $\mu_0H_\mathrm{ref}\approx\SI{229}{\milli\tesla}$ from the input to the output transducers $T_1$--$T_5$.
The subtraction of a reference signal excludes the electromagnetic crosstalk, but residual contributions around $f_\mathrm{FMR}\approx\SI{2.05}{\giga\hertz}$ remain. 
The measurements are compared to the SW excitation function $\rho_\mathrm{CSL}$ of the transducer effectively behaving like coplanar strip line (CSL).

The trMOKE images in Fig.~\ref{fig:single_tone}b–f visualize the corresponding wavefront propagation and deflection. 
At \SI{2.22}{\giga\hertz}, only the zero-order mode is observed at $T_3$, as $\lambda\ge d$ prevents diffraction. 
At \SI{2.38}{\giga\hertz}, the first-order wavefront is focused to $T_1$ and $T_5$, consistent with the peaks in $\Delta S_{21}$ in detection window $D_1$. 
At \SI{2.44}{\giga\hertz}, the wavefront is located between $D_1$ and $D_2$, resulting in a dip in the electrical signal. 
At \SI{2.50}{\giga\hertz}, the first-order mode shifts to $T_2$ and $T_4$, while a weak second-order contribution appears at $T_1$ and $T_5$.
This is clearly resolved in the trMOKE measurement, but barely detectable electrically.
The total input power splits up in the two modes while the first-order mode is excited more efficiently and therefore carries more power~\cite{Papp2017}. 
For frequencies above \SI{2.66}{\giga\hertz}, SWs with $\lambda\lesssim\SI{3}{\micro\meter}$ exhibit strong attenuation over the propagation path, and at the same time, $\rho_\mathrm{CSL}$ approaches its first minimum ($\approx\SI{2.7}{\giga\hertz}$).

The transmission measurements $|S_{21,H_x}|$ and $|S_{21,H_\mathrm{ref}}|$, a device characterization with varying input powers, and the full set of trMOKE measurements from \SIrange{2.22}{2.66}{\giga\hertz} are provided in Appendix~\ref{sec:A_Additional_YIG625}.
Except for the trMOKE measurements, the same RF characterization was performed around \SI{4.5}{\giga\hertz} and is as well provided in Appendix~\ref{sec:A_Additional_YIG625}.

\subsection{Two-Tone Excitation}
\label{ssec:Characterization_TwoTone}

The central functionality of the Rowland circle spectrometer is the spatial separation of SW wavefronts excited simultaneously at different frequencies at one external bias field. 
To demonstrate this capability, a two-tone trMOKE measurement was performed. 
The experiment is subject to two main constraints. 
First, the phase-locked periodic undersampling of the trMOKE setup requires both excitation frequencies to be integer multiples of the laser repetition rate (\SI{50}{\mega\hertz}) and the lock-in reference (\SI{100}{\kilo\hertz}). 
Second, the device operation is limited by the condition $\lambda_\mathrm{max}\leq 2\lambda_\mathrm{min}$ to avoid ambiguity due to overlapping diffraction orders. 
To satisfy these requirements, a reduced device geometry with \SI{100}{\nano\meter} YIG thickness, $R=\SI{36}{\micro\meter}$, and $d=\SI{8}{\micro\meter}$ was used. 

\begin{figure}[t!]
    \centering
    \includegraphics[width=\columnwidth]{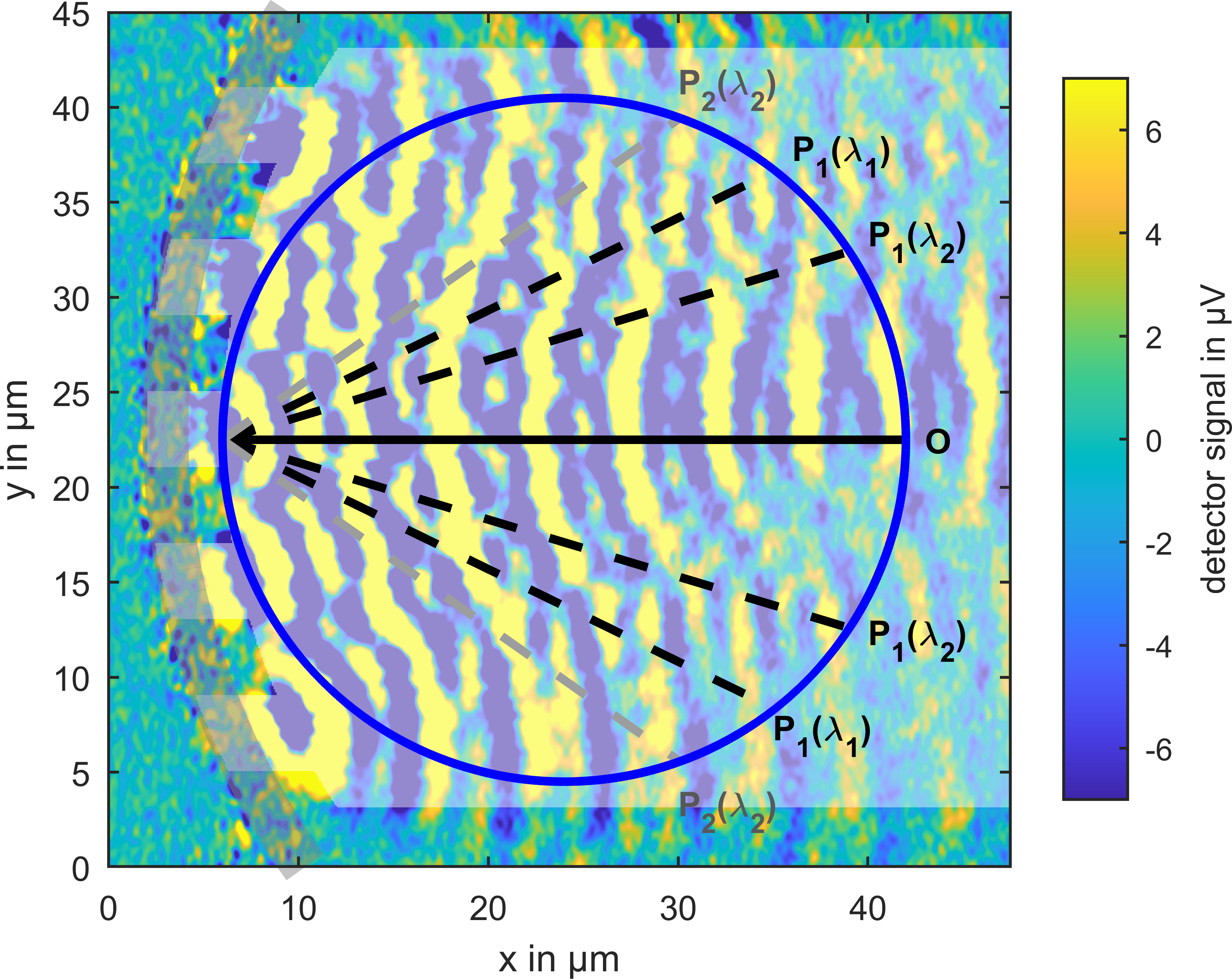}
    \caption{\textbf{Two-tone excitation of a Rowland circle spectrometer.} The measurement was conducted at a constant DC bias magnetic field of about \SI{220}{\milli\tesla} in out-of-plane direction. Simultaneous excitation at $f_1=\SI{2.10}{\giga\hertz}$ and $f_2=\SI{2.15}{\giga\hertz}$ results in SW with $\lambda_1=\SI{3.5}{\micro\meter}$ and $\lambda_2=\SI{2.3}{\micro\meter}$ and corresponding deflection angles $\alpha_1=\SI{26}{\degree}$ and $\alpha_2=\SI{17}{\degree}$. The analytically calculated deflection angles are indicated by the dashed lines. As adding signals in the frequency domain results in a beating pattern, the overall measurement shows a beating envelope with a periodicity of about \SI{6.7}{\micro\meter}. The spatial separation of wavefronts is only possible if the two frequency components are excited linearly in the grating.}
    \label{fig:TwoTone}
\end{figure}

The measurement was carried out at $\mu_0H\approx\SI{220}{\milli\tesla}$ using two phase-locked RF sources at $f_1=\SI{2.1001}{\giga\hertz}$ and $f_2=\SI{2.1501}{\giga\hertz}$.
The signals were added via a resistive power combiner and forwarded to the input transducer. 
The input frequencies correspond to SW wavelengths $\lambda_1\approx\SI{3.5}{\micro\meter}$ and $\lambda_2\approx\SI{2.3}{\micro\meter}$, yielding $\alpha_1=\SI{26}{\degree}$ and $\alpha_2=\SI{17}{\degree}$, respectively. 
Linear excitation of both frequency components is achieved by localized SW emission from the grating ridges, which is essential for spectrometer operation.
In Fig.\ref{fig:TwoTone}, the trMOKE colormap shows that the excitation of both wavelengths is possible, but a less clear wave pattern than for the single-tone excitation is observed.
The expected deflection points of the first two-order modes together with the second-order mode of $\lambda_1$ are added as reference.

The superposition of the two wave vectors leads to a spatial beating pattern described by
\begin{equation}
    \label{eq:Beating}
    \cos(k_1 x)+\cos(k_2 x)=2\cos\left(\frac{k_1+k_2}{2}x\right)\cos\left(\frac{k_1-k_2}{2}x\right),
\end{equation}
with $k_{1,2}=2\pi/\lambda_{1,2}$.
The second part in \eqref{eq:Beating} is equivalent to the carrier wave, while the first defines the beat envelope with a beat wave vector $\Delta k = |k_1-k_2|$.
The resulting trMOKE image is a snapshot of the propagating wavefronts, in which the modulation envelope with periodicity $\lambda_\mathrm{beat}=2\pi/\Delta k\approx\SI{6.7}{\micro\meter}$ is visible and matches the analytical model.
The beating dominates the patterns of the two diffracted wavefronts, but indicates the summation rather than the multiplication (mixing) of excited SW modes corresponding to the input frequencies.
In practice, the excitation amplitudes at varying SW wavelengths differ due to the frequency-dependent transducer excitation efficiency $\rho$. 
This imbalance was partially compensated by adjusting the sources' output powers to $P(f_1)=\SI{-1}{\dBm}$ and $P(f_2)=\SI{-2}{\dBm}$. 
Despite the experimental limitations, the measurement demonstrates linear excitation and the simultaneous propagation of multiple SW modes, extending previous demonstrations to multi-frequency operation.

\subsection{Electromagnetic Crosstalk}
\label{ssec:Characterization_Crosstalk}

\begin{figure}[ht!]
    \centering
    \includegraphics[width=\columnwidth]{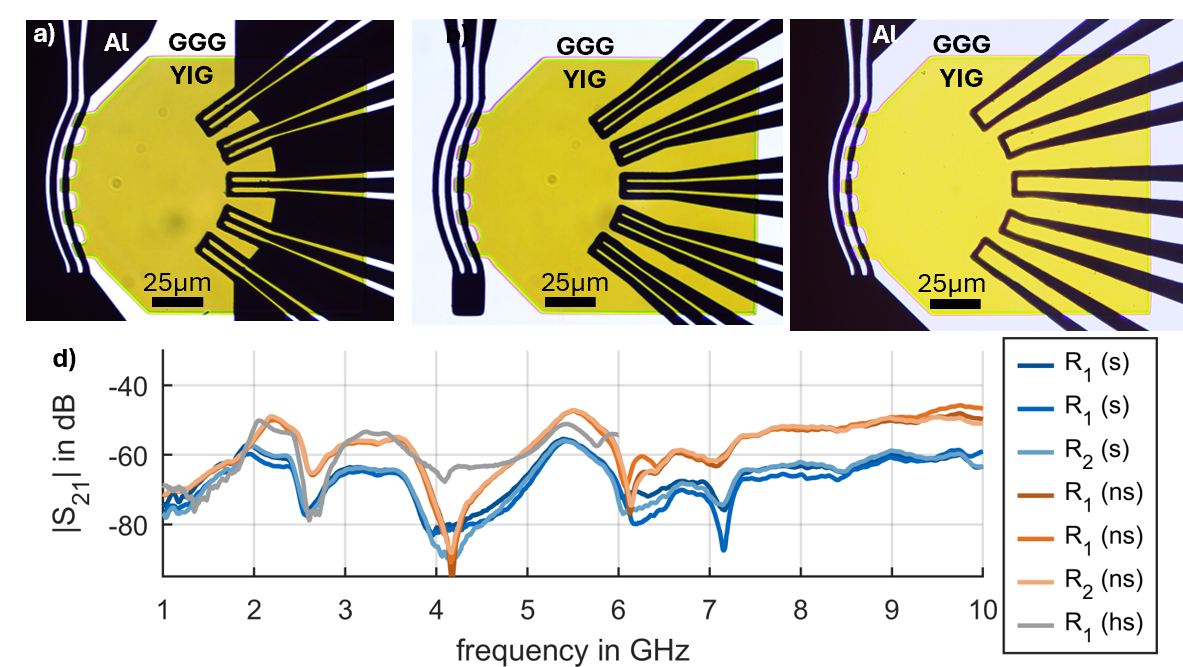}
    \caption{\textbf{Crosstalk evaluation of different Rowland circle variants.}  a)~The shielded (s) variant for the input/output transducers of the Rowland spectrometer with extended GND planes. b)~The non-shielded (ns) variant with GND lines only. Both pictures show fabricated Rowland circles with $R=\SI{75}{\micro\meter}$. c)~Calibrated transmission S-parameter measurements for two sizes of spectrometers ($R_1=\SI{75}{\micro\meter}$ and $R_2=\SI{90}{\micro\meter}$) with and without shielding. The gray graph shows the crosstalk of the half-shielded (hs) configuration from Fig.~\ref{fig:device_concept}a. Above \SI{2}{\giga\hertz}, the measurements show a systematic decrease of about \SI{10}{\decibel} in electromagnetic crosstalk when introducing wide GND planes.}
    \label{fig:crosstalk}
\end{figure}

Signal quality and detectability of the SW contribution at the local outputs depend strongly on the electromagnetic coupling between input and output transducers~\cite{Greil2023,Kohl2026}. 
To investigate this, several RF design variants were fabricated and characterized.
Two variants feature the shortened end of CPWs as output transducers, while the one used for the previously discussed measurements uses microstrip lines.
The first CPW variant extends the GND potential as far as possible without altering SW device functionality, forming GND planes on the input and output sides.
A through-light microscope image of this shielded variant is shown in Fig.~\ref{fig:crosstalk}a.
A direct comparison is made with a non-shielded variant that features GND lines rather than planes, as shown in Fig.~\ref{fig:crosstalk}b.
The third variant (Fig.~\ref{fig:crosstalk}c) is called half-shielded because extended GND planes are only feasible on the input side.
The non-shielded configuration is commonly used in magnonics~\cite{Vlaminck2010,Sushruth2020,Connelly2021,Lucassen2019,Kohl2026}, whereas the shielded one follows CPW design principles known from printed circuit board (PCB) technology~\cite{Simons2001}.
A general design goal for CPWs is to confine the electromagnetic field within the CPW gaps.
As a rule of thumb, the GND planes therefore should have a width of at least $3(w+2g)$ for a signal line of width $w$ and a gap width $g$.
For SW devices, this condition cannot be met near the excitation and detection areas, but can be partially satisfied in the tapered feed lines.

\begin{figure*}[ht!]
    \centering
    \includegraphics[width=\linewidth]{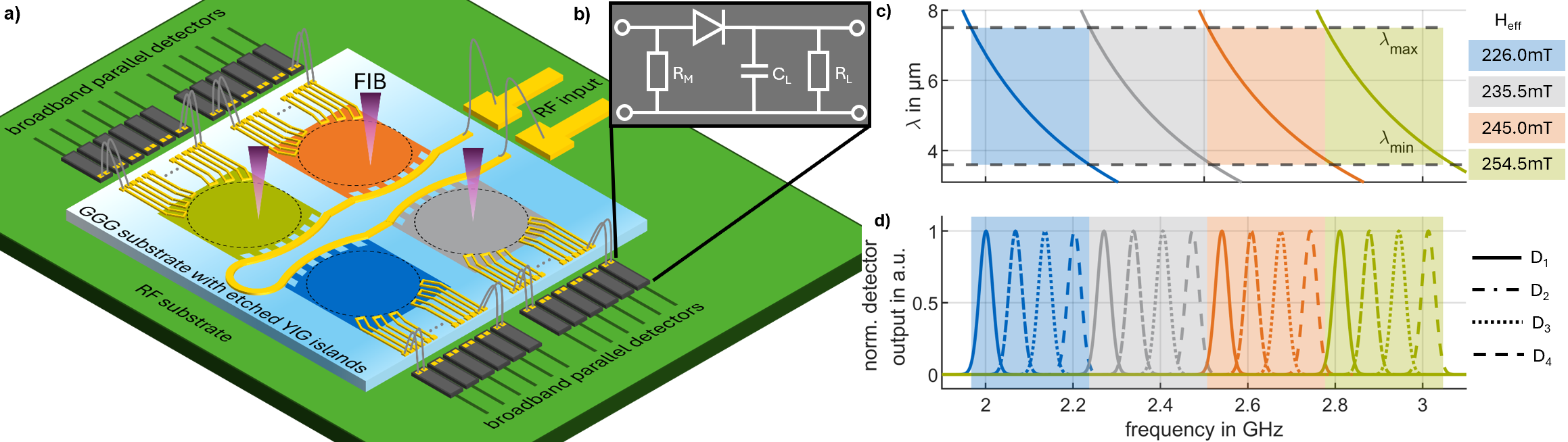}
    \caption{\textbf{Concept for a broadband and high-resolution magnonic spectrometer.} a)~Conceptual visualization of a parallelized SW-based spectrometer with four Rowland circles fabricated on a magnonic chip that is embedded in an RF substrate. The operational point of each spectrometer can be set, e.g., by focused-ion-beam irradiation at different dose levels. The spectrometers share a common input transducer, while multiple output transducers enable a parallel readout. b)~The output signals can be detected and converted to the electrical domain, e.g., via broadband RF power detectors placed on the RF substrate. c)~Exemplary dispersion relations for four spectrometers covering approximately the same wavelength range at different local effective fields $H_\mr{eff}$. d)~Assuming four local output transducers for equal detection spans $D_x$ of about \SI{50}{\mega\hertz}, it is possible to cover a total bandwidth of \SI{1}{\giga\hertz}.}
    \label{fig:DeviceIdeas}
\end{figure*}

Figure~\ref{fig:crosstalk}c compares calibrated transmission S-parameters of spectrometers with $R_1=\SI{75}{\micro\meter}$ and $R_2=\SI{90}{\micro\meter}$ for the layout variants in Fig.~\ref{fig:crosstalk}a-b and Fig.~\ref{fig:device_concept}a. 
For frequencies above \SI{2}{\giga\hertz}, the shielded CPW design yields about \SI{10}{\decibel} reduction in crosstalk, attributed to improved field confinement.
The half-shielded version behaves similarly to the unshielded variant and was used for the VNA characterization due to a more efficient signal pick-up.
The results highlight that magnonic device design calls for a co-optimization of SW and RF signal requirements. 
Feed lines and transducers with lateral dimensions in the range of tens to hundreds of micrometers may not be treated as purely lumped elements~\cite{Vanderveken2022,Erdelyi2025,Kohl2026}, but exhibit frequency-dependent RF behavior that critically influences device performance.
We identify the extension of GND planes as an effective strategy to suppress electromagnetic crosstalk, providing a transferable design principle for electrically interfaced magnonic devices.

\section{Device Integration and Application}
\label{sec:Realization_Application}

\subsection{Device Concept for Broadband and High-Resolution Spectrometers}
\label{ssec:Concept_MultiRC}

The Rowland circle geometry inherently defines a fixed operational range in the SW dispersion, resulting in a trade-off between spectral resolution and bandwidth. 
To relax this limitation, we envision a parallelized architecture in which multiple Rowland circles operate at different effective magnetic fields $H_\mathrm{eff}$. 
A conceptual implementation of this approach is shown in Fig.~\ref{fig:DeviceIdeas}a, where multiple spectrometers are integrated on a structured YIG/GGG chip embedded in a carrier substrate.
A shared input transducer and multiple parallel local output transducers are fabricated on the magnonic chip, while the signal detection using broadband RF power detectors (Fig.~\ref{fig:DeviceIdeas}b) is performed with conventional circuit elements on the carrier substrate.
Recent advances in microfabrication also indicate SW detection via a magnetoresistive scheme, as demonstrated in~\cite{Rossi2025}.
By exploiting the tunability and non-linearity of the SW dispersion, each spectrometer can be adjusted to a different operational frequency range while covering the same wavelength span, thereby enabling broadband operation with approximately constant frequency resolution.

The locally varying $H_\mathrm{eff}$ can be realized, e.g., by ion irradiation to modify $M_\mathrm{eff}$ in the YIG films~\cite{Kiechle2022,Kiechle2023,Bensmann2025,Naunheimer2026,Hoflich2023,Greil2025}, as indicated in Fig.~\ref{fig:DeviceIdeas}a.  
An example calculation of this concept is shown in Fig.~\ref{fig:DeviceIdeas}c for a Rowland circle with $R=\SI{75}{\micro\meter}$ and $d=\SI{16}{\micro\meter}$, illustrating that a total bandwidth of about \SI{1}{\giga\hertz} can be covered using four spectrometers with a static field offset of \SI{9.5}{\milli\tesla} to each other. 
For a realistic device design, we provide detailed, analytically calculated examples of three Rowland circles in Appendix~\ref{sec:A_MultiRC_Examples}.
Following these calculations, a realistic subdivision of each spectrometer's covered frequency range into four equidistant (in frequency) detection regions seems feasible (Fig.~\ref{fig:DeviceIdeas}d).
This results in detection bands with about \SI{50}{\mega\hertz} span and meet the bandwidths of communication channels in 5G-NR applications~\cite{3GPP38104} of up to \SI{100}{\mega\hertz}.

\subsection{Electrical Read-Out via Power Detection}
\label{ssec:Concept_Readout}

The proposed parallel architecture naturally motivates a simple and scalable readout strategy that uses, e.g., amplitude-based encoding, as the relevant spectral information is directly contained in the signal magnitude at the local outputs~\cite{IEEE802}. 
In terms of device integration, we propose broadband RF power detection as a simple, yet effective, and scalable interface between the magnonic device and conventional electronics. 
A conceptual realization of such a broadband power detector is illustrated in Fig.~\ref{fig:DeviceIdeas}b in the form of a half-wave rectifier. 
We implemented a first realization of such power detectors on a PCB using commercially available Schottky diodes (BAT63-02V~\cite{BAT63}) combined with a true-parallel multi-channel analog-to-digital converter (ADC,~\cite{ADS131M08}).
Details on the integration and characterization are provided in Appendix~\ref{sec:A_PowerDetector}.
Overall, we are able to demonstrate detection sensitivities down to approx.~\SI{-48}{\dBm} over \SIrange{0.1}{6}{\giga\hertz} in fully parallel measurements, but do not yet reach the detected SW signal levels using a VNA.

Nonetheless, the preliminary results confirm the detector concept's applicability and clearly indicate that further optimization should focus on the co-design of RF interfaces and magnonic structures. 
In particular, improved crosstalk mitigation, reduced interconnect lengths, and closer integration of detectors with the output transducers are expected to significantly enhance the signal-to-crosstalk ratio.
Under such conditions, compact, low-power readout schemes are compatible with the proposed multi-spectrometer architecture and can enable technology-relevant device implementations.

\subsection{Application Perspective: Low-Power Wake-Up Receivers}
\label{ssec:Usecase}

The combined concepts of spectral resolution and low-complexity detection suggest applications in RF front-ends for spectrum observation. 
A potential use case is clear channel assessment (CCA) within listen-before-talk (LBT) protocols~\cite{Zhang2015,Lagen2020,Ma2020,Baswade2021,Brauer2021}, in which spectral occupancy is evaluated to support dynamic resource allocation (channels or bandwidth) when sharing, e.g., licensed and unlicensed bands, as is the case in 5G-NR protocols~\cite{3GPP38104}. 
However, typical observation times of \SIrange{10}{25}{\micro\second}~\cite{ETSI_WAS} imply that the intrinsic fast response of SW devices does not directly improve protocol timing.

Instead, the combination of passive operation, low power consumption, and fast intrinsic propagation may be advantageous for energy-constrained front-end components, such as wake-up receivers~\cite{Hoglund2024}. 
With SW transit times on the order of \SIrange{30}{60}{\nano\second} (see Appendix~\ref{sec:A_ModelsAndCalcs}) and, therefore, the possibility of sub-\unit{\micro\second} threshold-based channel observation or link detection, the proposed device concept may enable fast and energy-efficient activation of subsequent circuitry. 
A corresponding detection scenario with multiple parallel detection bands is illustrated in Fig.~\ref{fig:DeviceIdeas}d, where up to \num{16} detection bands can be defined within a \SI{1}{\giga\hertz} bandwidth (see also Appendix~\ref{sec:A_MultiRC_Examples} for exemplary calculations). 
In this framework, the magnonic spectrometer may serve as a passive, frequency-selective preprocessing stage that operates without signal mixing, conversion, or digitization and is primarily used as a fast, low-power analog complement to conventional approaches.

\section{Conclusion}

In this work, we achieved a significant advancement in the realization of a magnonic Rowland circle spectrometer combined with a detailed electrical and magneto-optical characterization:  
The presented device integrates a structured YIG island, a curvilinear input transducer for coherent local SW excitation within the grating, and local electrical readout transducers into a single architecture. 
A key result is that the Rowland circle functionality can be implemented using a comparatively simple fabrication approach based on sputter deposition and wet-chemical etching, enabling concave YIG gratings with micrometer dimensions while preserving the magnetic film properties.
VNA and trMOKE characterization confirm that spatially separated SW wavefronts can be detected locally and that this detection correlates well with the expected diffraction behavior. 
The all-electrical readout demonstrates the transition from optical functionality investigation to electrically accessible device operation, which is crucial for device integration. 
Two-tone measurements show that simultaneous linear excitation of multiple SW wavelengths is feasible.
Finally, the demonstrated crosstalk reduction via extended GND planes represents a simple yet effective design strategy that can be readily adapted to a broad range of electrically interfaced magnonic devices without requiring significant modifications.

Despite the limitation of low SW signal amplitudes relative to the electromagnetic crosstalk, the results underline key advantages of SW-based signal processing. 
The operational point can be tuned via the bias field or magnetic material parameters, and the dispersive nature of SWs enables wavelength- and frequency-selective functionality in micrometer-scale structures at \unit{\giga\hertz}-frequencies.
This makes passive magnonic devices attractive for specialized analog signal-processing tasks at lowest power consumption.
The proposed parallelization of SW spectrometers serves as a case study that illustrates how the device concept can be extended to broadband operation while maintaining spectral resolution, thereby linking the device-level investigation to application-relevant scenarios.

Future work might focus on the co-optimization of electrical and magnonic design criteria, as addressed in, e.g.,~\cite{Erdelyi2025,Kohl2026}. 
Key steps include reducing crosstalk through RF co-design, improving the signal-to-crosstalk ratio, and integrating compact broadband detector concepts close to the magnonic device. 
Overall, the results indicate that a promising perspective lies in hybrid magnonic--electronic components being integrated into conventional RF systems. 
In such systems, SW devices may complement electronic circuitry by providing fast, passive, tunable, and low-power signal-processing functionality. 

\section*{Acknowledgements and Author Contributions}
The authors acknowledge funding from the European Union within
HORIZON-CL4-2021-DIGITAL-EMERGING-01 (No. 101070536, MandMEMS) and Deutsche Forschungsgemeinschaft (DFG, German Research Foundation) under project number 514146693.\\
J.G., F.N., and M.H. gratefully acknowledge the use of the cleanroom and lab facilities at Zentrales Elektronik- \& Informationstechnologielabor (ZEITlab) at TUM.\\

\textbf{J.G.} upgraded the wet-chemical YIG etching technology to a standardized process, fabricated samples, performed the trMOKE- and VNA-based device characterization, implemented the extension of GND planes for crosstalk reduction, and suggested the potential use case together with the multi-spectrometer device concepts. \textbf{M.H.} designed, implemented, and tested the parallel power detection read-out concept on the PCB level. \textbf{F.N.} performed micromagnetic simulations (Mumax) to back up the experimental parameter variations. \textbf{\'{A}.P.} initiated the adaptation of the Rowland circle concept to the SW domain and provided guidance for device design and evaluation. All authors contributed to conceptualization, discussion of results, and manuscript review.
\newpage
\bibliographystyle{IEEEtran}
\bibliography{NiceBib}

\begin{appendices}
\onecolumn


\section{Analytical Models and Calculations}
\label{sec:A_ModelsAndCalcs}
\subsection{Analytical Dispersion Relation and Group Velocity}
\label{ssec:A_Dispersion}

We used the SW dispersion relation developed by Kalinikos and Slavin~\cite{Kalinikos1986} to calculate
\begin{equation}
    \label{eq:dispersion}
    \omega = \sqrt{(\omega_0+\omega_\mr{M}\lambda_\mr{ex}k^2)(\omega_0+\omega_\mr{M}\lambda_\mr{ex}k^2+\omega_\mr{M}F)}
\end{equation}
with
\begin{equation}
    \label{eq:disp_F}
    F = P+\sin^2(\theta)\left(1-P(1+\cos^2(\phi))+\frac{\omega_\mr{M}P(1-P)\sin^2(\phi)}{\omega_0+\omega_\mr{M}\lambda_\mr{ex}k^2}\right),
\end{equation}

\begin{equation}
    \label{eq:disp_P}
    P = 1-\frac{1-e^{-tk}}{tk},
\end{equation}
$\omega_0=\gamma\mu_0H_\mr{eff}$, and $\omega_\mr{M}=\gamma\mu_0M_\mr{eff}$.
The effective field is $H_\mr{eff}=\mu_0H_\mr{ext}-M_\mr{eff}\cos(\theta)$.
The angles $\theta$ and $\phi$ define the direction of magnetization in a spherical coordinate system.
The material and geometrical parameters are the gyromagnetic ratio $\gamma$, the permeability of free space $\mu_0$, the exchange constant $\lambda_\mr{ex}$, and the thickness of the YIG film $t$.
The SW wavenumber $k$ is defined via the spin-wave wavelength $\lambda$ as $k=2\pi/\lambda$.
In the Forward-Volume (FV) configuration with $\theta=\SI{0}{\degree}$, i.e., $H_\mr{eff}$ pointint along the normal of the film surface, \eqref{eq:dispersion} simplifies to
\begin{equation}
    \label{eq:dispersion_OOP}
    \omega = \sqrt{(\omega_0+\omega_\mr{M}\lambda_\mr{ex}k^2)(\omega_0+\omega_\mr{M}\lambda_\mr{ex}k^2+\omega_\mr{M}P)}.
\end{equation}
For the calculation, we used $M_\mr{eff}=\SI{133}{\kilo\ampere\per\meter}$, $\lambda_{ex}=\SI{3.64e-12}{\joule\per\meter}$, and $t=\SIrange{100}{500}{\nano\meter}$, depending on the fabricated device.
The group velocity $v_\mr{g}$ is defined as
\begin{equation}
    \label{eq:GroupVelocity}
    v_\mr{g} = \frac{\partial\omega}{\partial k}
\end{equation}
and was used to calculate the transit time of the SWs through the Rowland circle.
From \eqref{eq:dispersion_OOP} and \eqref{eq:GroupVelocity}, one can extract group velocities and SW transit times $t_\mr{t}$ for the discussed frequency ranges and film thicknesses $t$ in this work, which are summarized in Tab.~\ref{tab:Velocity_Transition}.
The values correspond to the investigated magnonic Rowland circles at the two YIG film thicknesses $t=\SI{100}{\nano\meter}$ and $t=\SI{500}{\nano\meter}$, and at center frequencies $f_\mr{c}=\SI{2.4}{\giga\hertz}$ ($\mu_0H_\mr{ext}\approx\SI{246}{\milli\tesla}$) and $f_\mr{c}=\SI{4.65}{\giga\hertz}$ ($\mu_0H_\mr{ext}\approx\SI{327}{\milli\tesla}$). 
The SW transit time $t_\mr{t}=R/v_\mr{g}$ is calculated for a Rowland circle with $R=\SI{75}{\micro\meter}$.

\begin{table}[h!]
    \caption{Analytically calculated group velocities and transit times for the investigated Rowland spectrometers at two different YIG thicknesses and center frequencies.}
    \centering
    \label{tab:Velocity_Transition}
    \begin{tabular}{ccccc}
        \toprule
         & \makecell{$t=\SI{100}{\nano\meter}$\\ $f_\mr{c}=\SI{2.4}{\giga\hertz}$} & \makecell{$t=\SI{500}{\nano\meter}$\\ $f_\mr{c}=\SI{2.4}{\giga\hertz}$} & \makecell{$t=\SI{100}{\nano\meter}$\\ $f_\mr{c}=\SI{4.65}{\giga\hertz}$} & \makecell{$t=\SI{500}{\nano\meter}$\\ $f_\mr{c}=\SI{4.65}{\giga\hertz}$}\\
        \midrule
        $v_\mr{g,min}$ in \unit{\meter\per\second} & 490 & 1200 & 560 & 2050 \\
        $t_\mr{t,max}$ in \unit{\micro\second} & 0.15 & 0.063 & 0.13 & 0.037 \\
        $v_\mr{g,max}$ in \unit{\meter\per\second} & 550 & 2250 & 660 & 2720 \\
        $t_\mr{t,min}$ in \unit{\micro\second} & 0.13 & 0.033 & 0.11 & 0.028 \\
        \bottomrule
    \end{tabular}
\end{table}

\subsection{Transducer Excitation Function and Detection Spans}
\label{sec:A_rho_detection}

To extract the detection spans $D_{1,2}$, we first calculated the SW excitation spectrum (SWES) $\rho(k)$ for the transducer geometry.
Following the widely used approximation that the in-plane excitation field directly underneath a thin SW transducer carrying a uniform current density can be approximated by a constant rectangular function with width $w$, it is possible to approximate the SWES of a microstrip line (MSL) via~\cite{Sushruth2020,Connelly2021,Lucassen2019,Vlaminck2010}
\begin{equation}
    \label{eq:SWES_MSL}
    \rho_\mr{MSL}(k) \propto \frac{\sin(kw/2)}{kw/2},
\end{equation}
i.e., calculating the Fourier transform of the rectangular function in real space.
The excitation function is normalized.
The half-side grounded CPW transducer geometry used for the electrical characterization can be approximated by a coplanar strip line (CSL) or ground-signal (GS) line with a gap width $g$ by adding $\rho_\mr{MSL}(k)$ geometrically, and is written as\cite{Sushruth2020}
\begin{equation}
    \label{eq:SWES_CSL}
    \rho_\mr{CSL}(k) \propto 2\sin\left(\frac{k(w+g)}{2}\right)\rho_\mr{MSL}(k).
\end{equation}
For the calculation, a measured linewidth of $w=\SI{3}{\micro\meter}$ and gap $g=\SI{2}{\\micro\meter}$ are used.
Second, the same wavevector range $k$ is used to calculate analytically the expected deflection angles $\alpha$ in \eqref{eq:DeflectionAngle} for $n=1$ and $n=2$ as is visualized in Fig.~\ref{fig:device_concept}b-c in the main part.
From the fabricated Rowland circles, the minimum and maximum deflection angles $\alpha_\mr{T,min/max}$ at the corners of the transducers are calculated.
To define the detection regions, we assumed a point-like focusing of the SW wavefront.
In reality, the focal point is approximately on the order of the incident wavelength, which broadens the detection region.
However, a partial overlap of the deflected wavefront and the local output transducer results in very small signals that are barely detectable, thereby justifying the rather sharp definition of the detection areas $D_{1,2}$.
Finally, the same wavevector range $k$ was inserted to \eqref{eq:dispersion_OOP} using the experimental parameters (external bias field, film thickness) and the boundaries of $D_{1,2}$ were extracted corresponding to $\alpha_\mr{T,min/max}$ for every transducer.
As a result, $D_{1,2}$ contains information about the excitation efficiency of the SWs as well as their dispersive behavior with frequency at a constant bias field.


\section{Additional Data for half-shielded Rowland Circle With $R=\SI{75}{\micro\meter}$, $d=\SI{16}{\micro\meter}$}
\label{sec:A_Additional_YIG625}

\subsection{Electrical and Power-Dependent Characterization at \SI{2.45}{\giga\hertz}}
\label{ssec:A_YIG625_2p5GHz_electric}

\begin{figure*}[h]
    \centering
    \includegraphics[width=0.9\linewidth]{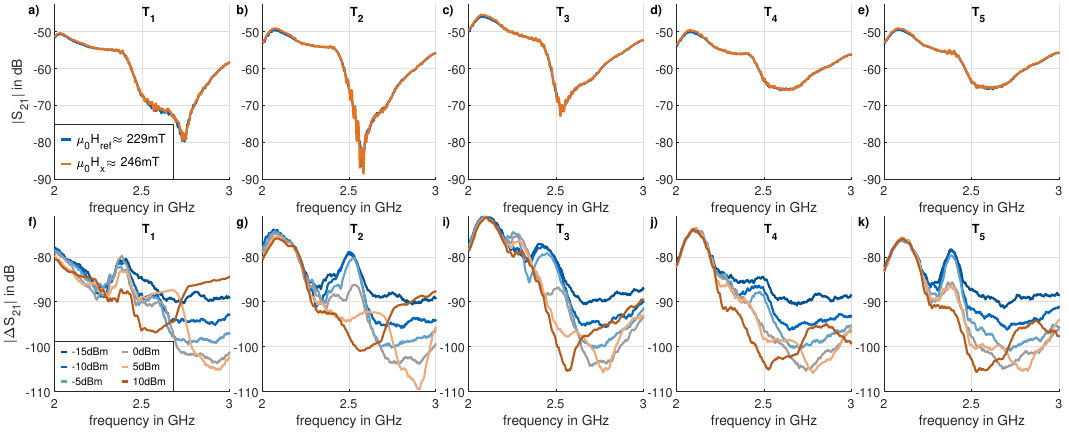}
    \caption{a)-e) $|S_{21}|$ for the five output transducers $T_1$--$T_5$ at $\mu_0H_\mr{ref} \approx \SI{220}{\milli\tesla}$ and $\mu_0H_x \approx \SI{246}{\milli\tesla}$. f)-k) Differential S-parameters $\Delta S_{21}$ for input powers from \SIrange{-15}{10}{\dBm}. Above about \SI{-10}{\decibel} the SWs are excited non-linearly, which marks the region of the $\mathrm{1dB}$ compression point.}
    \label{fig:S21_Abs_Pow_537}
\end{figure*}

The magnitude of the forward transmission $|S_{21}|$ for all outputs is shown in Fig.~\ref{fig:S21_Abs_Pow_537}a)-e) for the reference measurement at $\mu_0H_\mr{ref} \approx \SI{220}{\milli\tesla}$ and the actual measurement at $\mu_0H_x \approx \SI{246}{\milli\tesla}$.
In the presented frequency span $T_{4,5}$, a less pronounced dip is shown at around \SI{2.5}{\giga\hertz}, which was observed as a systematic difference between $T_{1,2,3}$ and $T_{4,5}$ in other Rowland circles with the same geometry and on the same chip.
We attribute this to slight deviations in fabrication, as, e.g., the covering of the etched YIG edge depends on the local quality of the evaporated metallization (the sample was rotated during evaporation).
In Fig.~\ref{fig:S21_Abs_Pow_537}f)-k), the power-dependent analysis of the differential S-parameters $|\Delta S_{21}|$ is shown for input powers from \SIrange{-15}{10}{\dBm}.
For input powers above about \SI{-10}{\dBm}, the SWs start getting excited in the non-linear regime and as a result the amplitude of the transmitted SW signals drops~\cite {Erdelyi2026}.
This marks the region of the device's 1dB compression point, $P_\mr{1dB}$.

\begin{figure*}[h]
    \centering
    \includegraphics[width=0.8\linewidth]{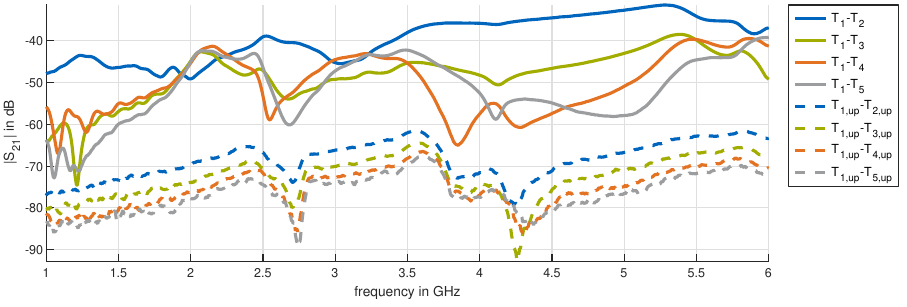}
    \caption{Solid lines: Electromagnetic crosstalk between the output transducers of a Rowland circle with $R=\SI{75}{\micro\meter}$ that is half-shielded. Dashed lines: Reference measurement for the electromagnetic crosstalk between the GSG probes at the same lateral position but lifted about \SI{5}{\milli\meter} above the YIG chip surface. In contact, the crosstalk is about \SI{30}{\decibel} higher, which indicates that the direct GSG-probe crosstalk plays a negligible role in the measurement.}
    \label{fig:YIG625_b2_OutputCrosstalk}
\end{figure*}

With regard to a potential downscaling and integration of the presented Rowland spectrometer, we also characterized the electromagnetic crosstalk between the output transducers.
The measurements are shown in Fig.~\ref{fig:YIG625_b2_OutputCrosstalk}, where the solid lines represent the state in which the two GSG probes are in contact with the landing pads of the output transducers.
The next neighbors $T_1$ and $T_2$ show systematically the highest crosstalk, while the increased distance from $T_1$ to $T_{3,4,5}$ lowers the crosstalk about \SIrange{0}{20}{\decibel}, depending on the frequency.
As a reference for the crosstalk between the GSG probes at the same distance, the probes were lifted about \SI{5}{\milli\meter} above the sample surface, and the same measurement was repeated, as indicated by the dashed lines.
Compared with the measurement in contact, the corresponding crosstalk is about \SI{30}{\decibel} lower and, therefore, negligible.

\subsection{TrMOKE Characterization at \SI{2.45}{\giga\hertz}}
\label{ssec:A_YIG625_2p5GHz_trMOKE}

\begin{figure*}[h!]
    \centering
    \includegraphics[width=0.9\linewidth]{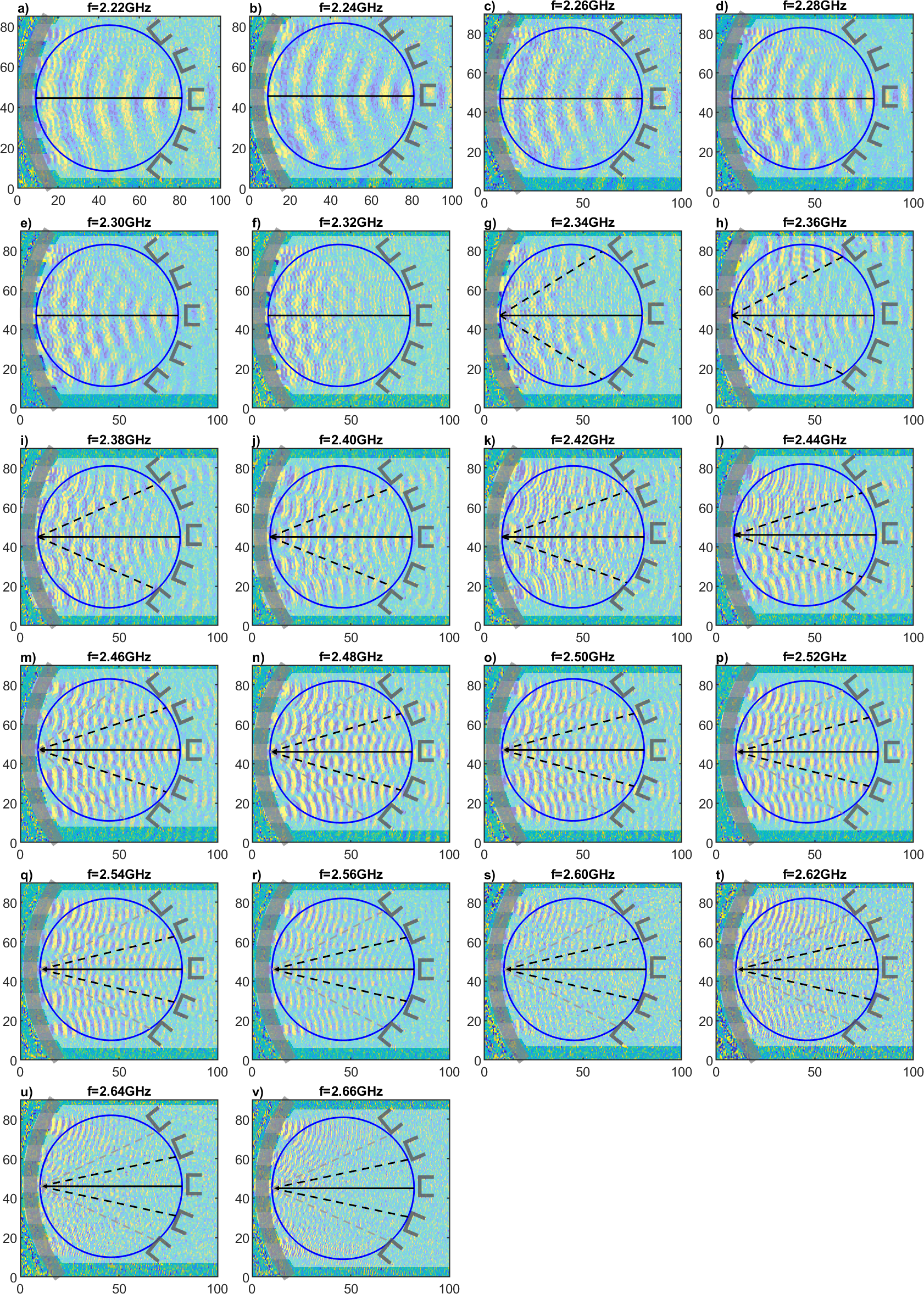}
    \caption{TrMOKE measurements of the Rowland circle (blue) with $R_1=\SI{75}{\micro\meter}$ and $d=\SI{16}{\micro\meter}$ at a bias field $\mu_0H_\mr{ext}\approx\SI{246}{\milli\tesla}$ from \SIrange{2.22}{2.66}{\giga\hertz} in steps of \SI{0.02}{\giga\hertz}. The concave grating and input/output transducers are overlaid on the measurements. The black and gray dashed lines show the deflection angle $\alpha_{1,2}$ for the first and second order diffraction mode in the grating.}
    \label{fig:YIG625_t3_trMOKE}
\end{figure*}

In Fig.~\ref{fig:YIG625_t3_trMOKE}, a set of trMOKE measurements of the half-shielded Rowland circle with $R_1=\SI{75}{\micro\meter}$ and $d=\SI{16}{\micro\meter}$ are shown that were conducted at a bias field $\mu_0H_\mr{ext}\approx\SI{246}{\milli\tesla}$ and for exctiation frequencies from \SIrange{2.22}{2.66}{\giga\hertz} in steps of \SI{0.02}{\giga\hertz}.
The nominal output power of the RF source used for excitation was \SI{0}{\dBm} for Fig.~\ref{fig:YIG625_t3_trMOKE}a-s and \SI{3}{\dBm}, \SI{5}{\dBm}, and \SI{7}{\dBm} for Fig.~\ref{fig:YIG625_t3_trMOKE}t-v, respectively (Figure at end of Appendix for better formating).
The available power at the SW transducer is estimated to be at least \SI{10}{\decibel} below the source's output power due to lossy cables, adapters, feed lines on the carrier PCB, long bond wires, and the omitted impedance matching.
The concave grating is overlaid on the colormap as a semi-transparent white array.
The curved transducer and the tips of the local output transducers are overlaid on the colormap as semi-transparent gray graphs.
The blue circle denotes the Rowland circle with a diameter of $R_1=\SI{75}{\micro\meter}$.
The black and gray dashed lines show the deflection angle $\alpha_{1,2}$ for the first and second order diffraction mode in the grating.
The colormap has the same scale ($[-15,15]$, detector signal in \unit{\micro\volt}) for Fig.~\ref{fig:YIG625_t3_trMOKE}a-r and a rescaled one for Fig.~\ref{fig:YIG625_t3_trMOKE}s-v ($[-10,10]$).

Below \SI{2.30}{\giga\hertz}, only the zero-order focusing effect of the curvature is visible because $\lambda\ge d$.
However, on the lower side, a slight focusing toward $T_5$ can be seen.
From \SIrange{2.34}{2.38}{\giga\hertz}, the first-order diffraction mode is deflected to the transducers $T_{1,5}$.
From \SIrange{2.40}{2.42}{\giga\hertz}, the wavefronts are deflected between the transducers, and no output signal can be detected.
From \SI{2.44}{\giga\hertz} on the first-order mode is deflected to the output transducers $T_{2,4}$, while the second-order mode is visible from \SI{2.46}{\giga\hertz} onwards.
Starting at around \SI{2.56}{\giga\hertz}, the excitation efficiency of the transducer drops, and the output power of the source must be increased to maintain a detectable signal.
At the same time, the group velocity of the short-wavelength SWs decreases, which results in an effectively increased damping per wavelength.
Therefore, the achievable propagation distance decreases to a level where no more (detectable) SWs reach the output transducers.

\subsection{Electrical and Power-Dependent Characterization at \SI{4.65}{\giga\hertz}}
\label{ssec:A_YIG625_4p5GHz_electric}

\begin{figure*}[h]
    \centering
    \includegraphics[width=0.9\linewidth]{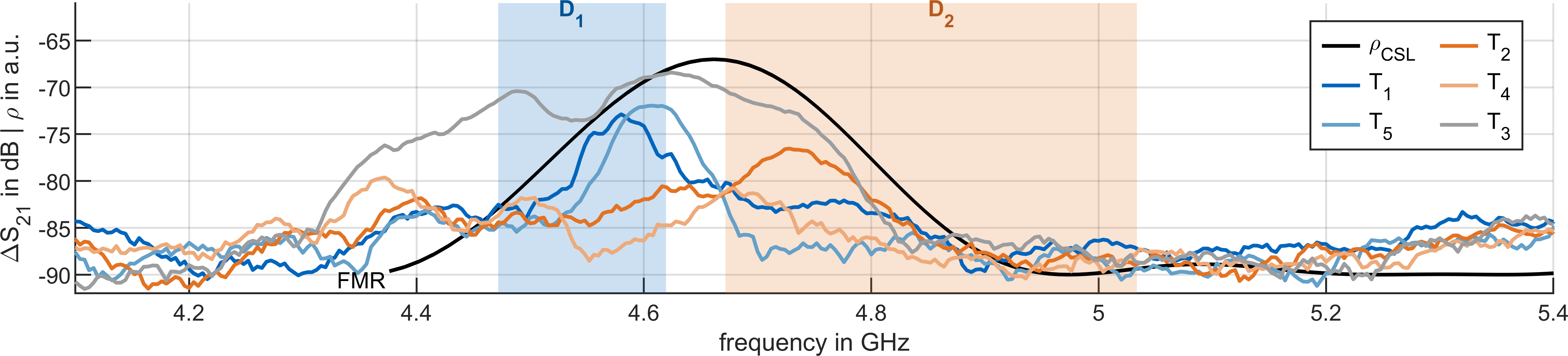}
    \caption{a) Calibrated  differential S-parameter measurements $|\Delta S_{21}|$ for a concave grating with $R=\SI{75}{\micro\meter}$ and $d=\SI{16}{\micro\meter}$ at a constant bias field $\mu_0H_x \approx \SI{327}{\milli\tesla}$. The measured detection peaks for the symmetrical output transducers $T_1/T_2$ and $T_4/T_5$ are located in the analytically calculated detection regions $D_1$ and $D_2$. The zero-order focusing point $O$ serves as the reference signal at $T_3$. The excitation function for the used input transducer geometry can be approximated by a coplanar strip line and is overlaid on the measurements. Due to lower electromagnetic crosstalk in the presented frequency span, the spurious signals are small compared to the SW signal.}
    \label{fig:single_tone_4GHz}
\end{figure*}

An additional characterization of the Rowland circle was performed at a bias field of $\mu_0H \approx \SI{327}{\milli\tesla}$ and a reference field of $\mu_0H_\mr{ref} \approx \SI{300}{\milli\tesla}$.
The corresponding differential S-parameters $|\Delta S_{21}|$ of the five consecutive output transducers together with the analytically calculated transducer excitation function $\rho_\mr{CSL}$ are presented in Fig.~\ref{fig:single_tone_4GHz}.
Due to a flatter region in the electromagnetic crosstalk paired with an increased absolute SW excitation efficiency of the transducer in the frequency range around $f_\mr{c}\approx\SI{4.65}{\giga\hertz}$, the field-dependent spurious signal contributions are negligible.
As a result, the differential SW signal is about \SI{20}{\decibel} above the noise floor of the measurement.
While with this measurement $\rho_\mr{CSL}$ agrees well with the zero-order focusing reference transducer $T_3$, the detection amplitude of $T_4$ still remains comparatively small.

\begin{figure*}[h]
    \centering
    \includegraphics[width=0.9\linewidth]{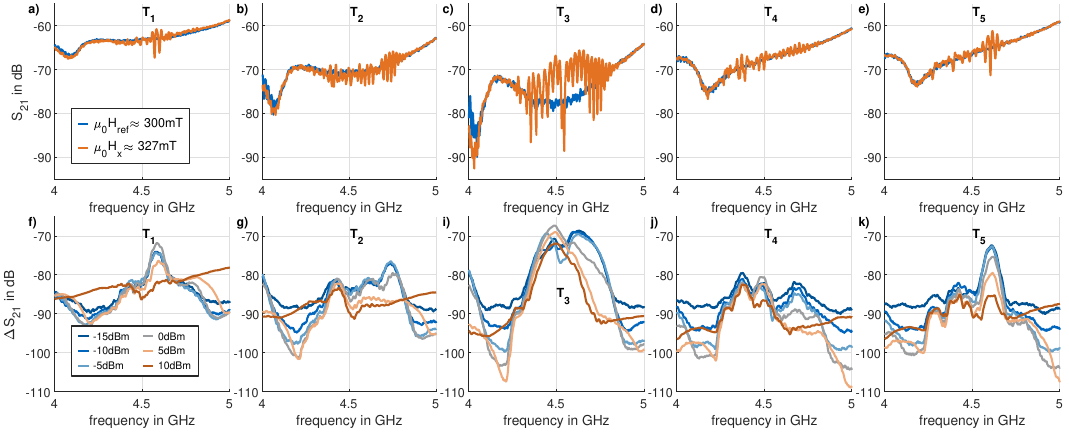}
    \caption{a)-e) $|S_{21}|$ for the five output transducers $T_1$--$T_5$ at $\mu_0H_\mr{ref} \approx \SI{300}{\milli\tesla}$ and $\mu_0H_x \approx \SI{327}{\milli\tesla}$. The flat electromagnetic crosstalk enables clearer SW signal extraction. f)-k) Differential S-parameters $\Delta S_{21}$ for input powers from \SIrange{-15}{10}{\dBm}. Above about \SI{-10}{\decibel}, the SWs are excited non-linearly, which marks the region of the $\mathrm{1dB}$ compression point $P_\mathrm{1dB}$.}
    \label{fig:S21_Abs_Pow_710}
\end{figure*}

The magnitude of the forward transmission $|S_{21}|$ for all transducers is shown in Fig.~\ref{fig:S21_Abs_Pow_710}a)-e) for the reference measurement at $\mu_0H_\mr{ref} \approx \SI{300}{\milli\tesla}$ and the actual measurement at $\mu_0H_x \approx \SI{327}{\milli\tesla}$.
The comparison of the signals shows that the two outermost transducers systematically exhibit the highest crosstalk, as their tapered feed lines span the largest loop that can be penetrated by the near-field electromagnetic field, thereby inducing the crosstalk.
$T_2$ and $T_4$ show thefore about \SIrange{5}{10}{\decibel} less crosstalk, while $T_3$ has the shortest tapered feed lines and shows about \SIrange{5}{15}{\decibel} less crosstalk.
In Fig.~\ref{fig:S21_Abs_Pow_710}f)-k), the power-dependent analysis of the differential S-parameters $|\Delta S_{21}|$ is shown for input powers from \SIrange{-15}{10}{\dBm}.
The same general tendency as in Fig.~\ref{fig:S21_Abs_Pow_537}f)-k) can be observed, and above about \SI{-10}{\dBm} the SWs start getting excited in the non-linear regime.

\section{Additional Data for Crosstalk Characterization of Shielded and Unshielded Rowland Circles}
\label{sec:A_Additional_YIG622}

\begin{figure*}[h!]
    \centering
    \includegraphics[width=0.9\linewidth]{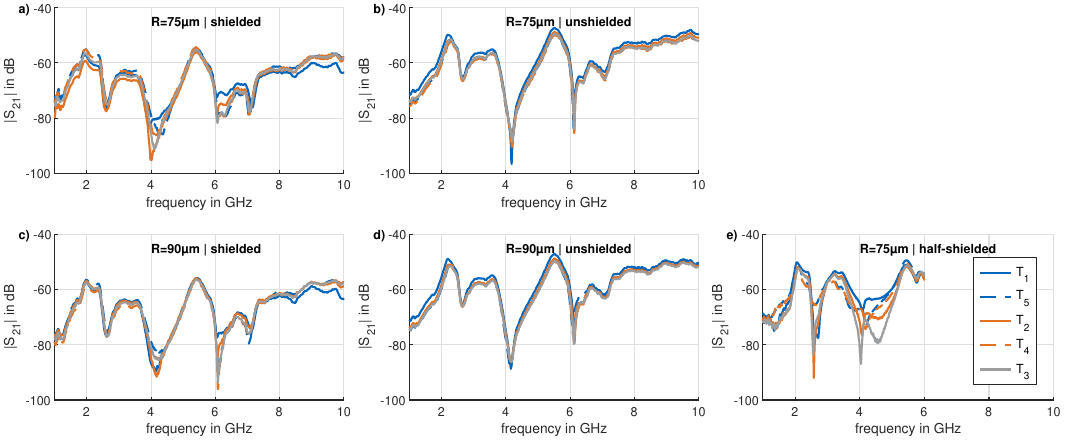}
    \caption{Characterization of the electromagnetic crosstalk between the output transducers for shielded variants with extended GND planes in a) and c), which are compared to the unshielded variants of the same grating size in b) and d). Above \SI{2}{\giga\hertz}, the shielded variants show about \SI{10}{\decibel} less crosstalk. The half-shielded variant in e), used for local electrical detection in the main part, behaves similarly to the unshielded variants.}
    \label{fig:CrosstalkComparison_YIG622_YIG625}
\end{figure*}

The forward transmission $|S_{21}|$ (crosstalk) of all electrically characterized Rowland circles discussed in this work is shown in Fig.~\ref{fig:CrosstalkComparison_YIG622_YIG625} for all output transducers.
The shielded variants with $R=\SI{75}{\micro\meter}$ \& $d=\SI{16}{\micro\meter}$ and $R=\SI{90}{\micro\meter}$ \& $d=\SI{20}{\micro\meter}$ in Fig.~\ref{fig:CrosstalkComparison_YIG622_YIG625}a) and Fig.~\ref{fig:CrosstalkComparison_YIG622_YIG625}c) have consistently about \SI{10}{\decibel} less crosstalk than the unshielded variants of same size in Fig.~\ref{fig:CrosstalkComparison_YIG622_YIG625}b) and Fig.~\ref{fig:CrosstalkComparison_YIG622_YIG625}d).
For comparison, we plot the same characterization performed for the half-shielded Rowland circle with $R=\SI{75}{\micro\meter}$ \& $d=\SI{16}{\micro\meter}$ in Fig.~\ref{fig:CrosstalkComparison_YIG622_YIG625}e).
Further investigations are needed to reduce crosstalk to an acceptable level relative to the achievable SW signal and to make it as flat as possible across a broad frequency span.
Nonetheless, the simple layout adaptation demonstrates the potential for fine-tuning RF chip design alongside pure-magnonic device design.

\begin{figure*}[h!]
    \centering
    \includegraphics[width=0.9\linewidth]{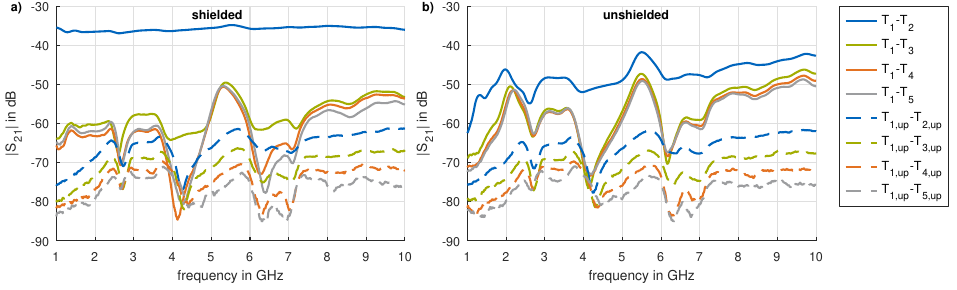}
    \caption{Comparison between the crosstalk of a) shielded and b) unshielded Rowland circles with $R=\SI{75}{\micro\meter}$ \& $d=\SI{16}{\micro\meter}$. The dashed lines indicate reference measurements taken at the positions of the contact pads, with the GSG probes elevated by about \SI{5}{\milli\meter} above the sample surface. The next-nearest neighbors exhibit the strongest crosstalk in both variants, while the nearby shorted GND planes in the shielded spectrometer enable strong yet flat coupling between the channels. In general, the shielding reduces the crosstalk between the outputs by \SIrange{2}{5}{\decibel}.}
    \label{fig:YIG622_OutputCrosstalks}
\end{figure*}

We performed the same crosstalk evaluation between the output transducers as shown in Fig.~\ref{fig:YIG625_b2_OutputCrosstalk} for the shielded and unshielded Rowland circle of the same size ($R=\SI{75}{\micro\meter}$ \& $d=\SI{16}{\micro\meter}$), which are presented in Fig.~\ref{fig:YIG622_OutputCrosstalks}.
For the shielded version in Fig.~\ref{fig:YIG622_OutputCrosstalks}, the most significant difference is the crosstalk between $T_1$ and $T_2$ as those next-nearest neighbors are shorted and, thus, the crosstalk is comparatively high yet flat across the tested frequency range.
Besides this, the measurement shows in principle the same behavior as the half-shielded version.
The reference measurements with lifted GSG probes show about \SIrange{10}{20}{\decibel} less crosstalk than when the probes are in contact with the metallization.
A similar picture can be observed for the unshieled version shown in Fig.~\ref{fig:YIG622_OutputCrosstalks}, whereas the crosstalk between $T_1$ and $T_2$ is still significantly higher than to $T_{3,4,5}$, but follows the same trend.
In general, the crosstalk between the outputs of the shielded version is about \SIrange{2}{5}{\decibel} less than in the unshielded case, even if all GND lines/planes are shorted.
This trend clearly indicates that the local electromagnetic field confinement in the gaps of the CPW/CPS transducers plays a significant role in the local guiding of the electromagnetic signal and should be considered alongside the magnonic design.

\section{Implementation and Characterization of a Parallel Power-Detector Circuit}
\label{sec:A_PowerDetector}

\begin{figure*}[h!]
    \centering
    \includegraphics[width=0.9\linewidth]{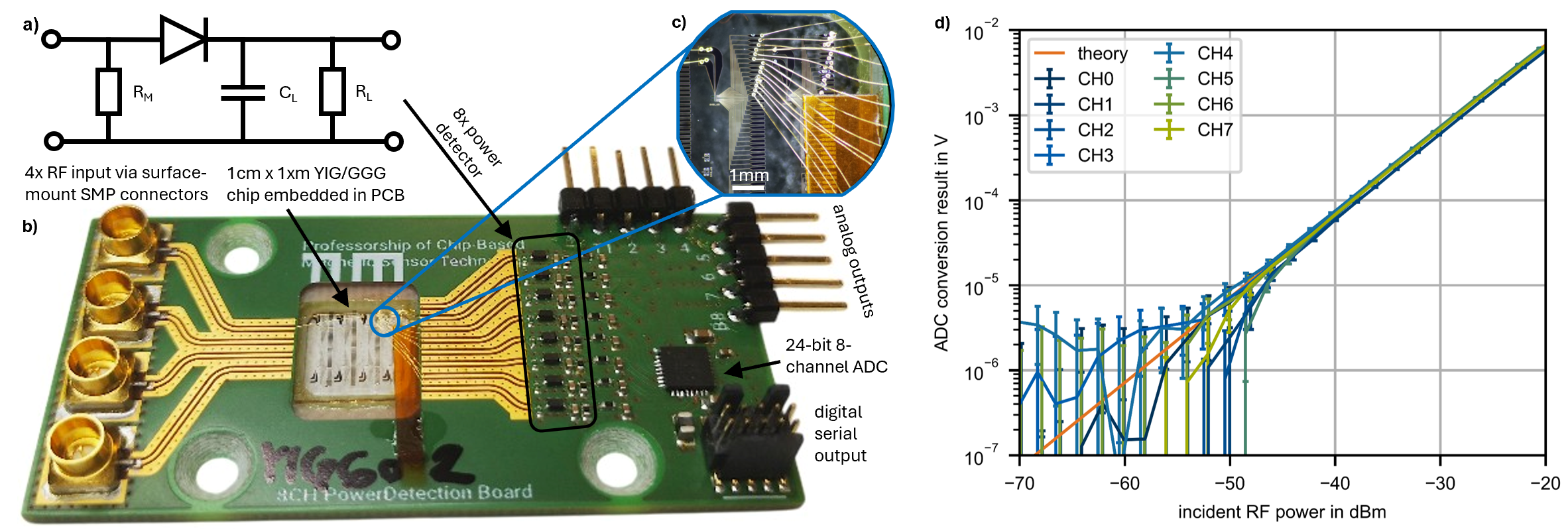}
    \caption{a) Schematic of an RF power detector consisting of a matching resistor $R_\mr{M}$, a Schottky diode, a load capacitor $C_\mr{L}$, and a load resistor $R_\mr{L}$. b) Photograph of a PCB with \num{4} input channels, a milled pocket for embedding the \SI{1}{\centi\meter}~$\times$~\SI{1}{\centi\meter} YIG/GGG chip featuring several Rowland circles with feed lines and bond pads, \num{8} parallel power detectors, and a 24-bit 8-channel true-parallel analog-to-digital converter (ADC). The ADC can be read out via a serial interface for digital outputs. The PCB also provides access to the rectified analog signals via the pin headers. c) Microscope image of a bonded Rowland circle via gold ball bonds. d) Characterization of the eight implemented parallel power detectors wire-bonded to a test coupon holding \num{4} CPW traces in the middle of the PCB. Reliable power detection down to about \SI{-48}{\dBm} is possible with the presented circuit, which can be further optimized.}
    \label{fig:Realization_PCB}
\end{figure*}

The parallel power detector concept utilizing a half-wave rectifier based on a Schottky diode (Fig.~\ref{fig:Realization_PCB}a) was implemented and experimentally validated on a dedicated printed circuit board, shown in Fig.~\ref{fig:Realization_PCB}b. 
The realized system integrates up to four SMP inputs and eight identical Schottky-diode-based detector channels, along with the YIG/GGG chip, enabling fully parallel RF power readout.
The YIG/GGG chip is embedded and glued into a milled pocket in the PCB and wire-bonded via \SI{25}{\micro\meter} gold ball bonds to the eight output channels, as shown in Fig.~\ref{fig:Realization_PCB}c.
Each detector follows a simple envelope-detection topology consisting of a \SI{50}{\ohm} matching resistor, a Schottky diode for rectification (BAT63-02V~\cite{BAT63}), and an RC low-pass network ($C_\mr{L}=\SI{1}{\nano\farad}$ and $R_\mr{L}=\SI{100}{\kilo\ohm}$). 
The resulting DC voltages are acquired simultaneously via a multi-channel, high-resolution ADC~\cite{ADS131M08} integrated directly on the PCB, enabling synchronized digital readout without external measurement instrumentation. 
Using a microcontroller, the ADC is read out by a computer via a serial interface.

Before milling the pocket for embedding the magnonic chip, the output channels were characterized in groups of 4 by bonding the input to the output channels via a coupon with 4 parallel CPW traces. 
In Fig.~\ref{fig:Realization_PCB}d, the digitized voltage output of the eight output channels at power levels lower than \SI{-20}{\dBm} is shown at a measurement frequency of \SI{2.45}{\giga\hertz}. 
The input power refers to the calibrated power on the SMA-to-SMP adaptor cable used in the experiment, resulting in a slightly lower incident power at the detectors due to small losses in the adapter, the CPW traces on the PCB, and bond-wire mismatches. 
According to~\cite{Thumm-HF-MST}, the DC output voltage of the Schottky power detector is directly proportional to the input power, which lines up with the measurement result in Fig.~\ref{fig:Realization_PCB}d for sufficiently high powers. 
The combination of this dependence with the logarithmic scale typically used for RF powers requires measuring voltages on a logarithmic scale.
At low power levels, the ambient noise floor and the ADC's accuracy fundamentally limit the overall power detection capabilities. 
For the presented concept board, the sensitivity limit is roughly \SI{-48}{\dBm}, which is about \SIrange{20}{25}{\decibel} above the level needed for signal detection in the presented Rowland spectrometers.
Nonetheless, the investigation shows that, using comparatively simple and not yet fully optimized detection circuits, it is possible to approach an integrated, local electrical read-out scheme.
Further optimization of component characteristics, physical circuit integration, and magnonic device design is required to close the remaining gap, which we estimate is feasible using standard commercial components. 

\newpage


\section{Examples for Choosing Operational Points and Detection Spans for Magnonic Spectrometers}
\label{sec:A_MultiRC_Examples}

\begin{figure*}[t!]
    \centering
    \includegraphics[width=0.9\linewidth]{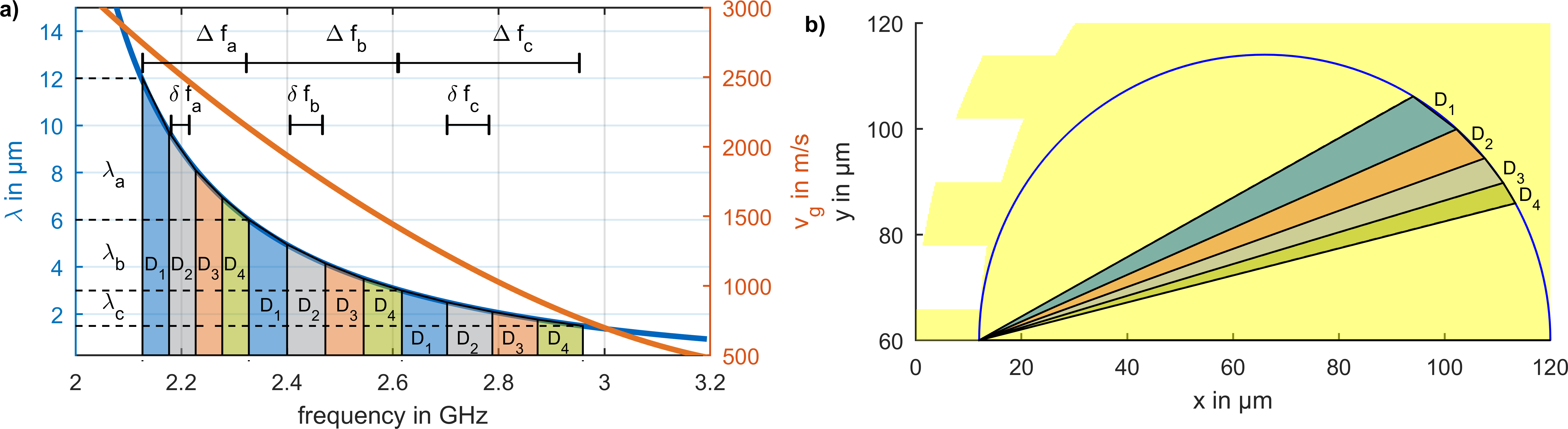}
    \caption{Exemplary calculation for possible working regimes of SW-based spectrometers. a) Analytically calculated dispersion relation for $M_\mr{eff}=\SI{133}{\kilo\ampere\per\meter}$, $\mu_0H_\mr{ext}=\SI{236}{\milli\tesla}$ and $d=\SI{500}{\nano\meter}$. Three wavelength ranges $\lambda_a=[12,6]~\unit{\micro\meter}$, $\lambda_b=[6,3]~\unit{\micro\meter}$, $\lambda_c=[3,1.5]~\unit{\micro\meter}$ correspond to three frequency spans $\Delta f_a=\SI{201}{\mega\hertz}$, $\Delta f_b=\SI{289}{\mega\hertz}$, and $\Delta f_c=\SI{342}{\mega\hertz}$ that can be covered by a single Rowland circle. The frequency spans are split up into four equally broad (in frequency) detection bands $\delta f_a=\SI{50.3}{\mega\hertz}$, $\delta f_b=\SI{72.3}{\mega\hertz}$, and $\delta f_b=\SI{86.1}{\mega\hertz}$.  b) Schematic drawing of a Rowland circle with $R_a=\SI{120}{\micro\meter}$, $d_a=\SI{24}{\micro\meter}$ that is able to distinguish \num{4} detection bands (channels) with a bandwidth of $\delta f_a=\SI{50.3}{\mega\hertz}$ within $\Delta f_a$. The lengths of the corresponding output transducers along the Rowland circle (blue line) are \SIrange{4.4}{11}{\micro\meter}. When scaling $R$ and $d$ with the same factor, the covered range of $\alpha$ remains the same, while the arc length of the transducers scales linearly.}
    \label{fig:MultiRC_estimation}
\end{figure*}

In this section, we present an exemplary discussion of possible specific realizations of Rowland circle devices.
The parameters are chosen to remain in a technologically feasible range.
The starting point is the range of wavelengths that should be used in the devices and is extracted from \eqref{eq:dispersion_OOP} for $M_\mr{eff}=\SI{133}{\kilo\ampere\per\meter}$, $\mu_0H_\mr{ext}=\SI{236}{\milli\tesla}$ and $d=\SI{500}{\nano\meter}$.
The dispersion is plotted in Fig.~\ref{fig:MultiRC_estimation}a) on the left y-axis.
We can identify three operational ranges that fulfill $\lambda_\mr{max}\le 2\lambda_\mr{min}$ to ensure an unambigous frequency detection: $\lambda_a=[12,6]~\unit{\micro\meter}$, $\lambda_b=[6,3]~\unit{\micro\meter}$, $\lambda_c=[3,1.5]~\unit{\micro\meter}$. 
This allows to fabricate Rowland circles in three different technologically feasible sizes: $R_a=\SI{120}{\micro\meter}$, $d_a=\SI{24}{\micro\meter}$, $R_b=\SI{60}{\micro\meter}$, $d_b=\SI{12}{\micro\meter}$, $R_c=\SI{30}{\micro\meter}$, $d_c=\SI{6}{\micro\meter}$, whereas $d$ is as well defined by $\lambda$.
To maintain comparability across sizes, we fix $R$ so that at least \num{5} grating ridges define the concave grating, as it was demonstrated in the main part.
A somewhat free design parameter is $R$ and the arc length of the grating compared to $d$, as the absolute arclength between the separated wavefront along the Rowland circle line increases with $R$ at constant $d$, but so does the propagation distance, which naturally limits the ratio between $R$ and $d$.
By scaling $R$ and $d$ together, we fix this ratio and implicitly assume that the effective damping per SW wavelength changes approximately with the same ratio.
From Fig.\ref{fig:MultiRC_estimation}a, we identify the corresponding three frequency spans $\Delta f_a=\SI{201}{\mega\hertz}$, $\Delta f_b=\SI{289}{\mega\hertz}$, and $\Delta f_c=\SI{342}{\mega\hertz}$ that can be covered by a single Rowlan circle, and separated them into \num{4} equally broad detection bands $\delta f_a=\SI{50.3}{\mega\hertz}$, $\delta f_b=\SI{72.3}{\mega\hertz}$, and $\delta f_c=\SI{86.1}{\mega\hertz}$.
Inserting the wavelenghts corresponding to the frequenices at the borders of the detection bands provides the deflection angles $\alpha_1=\SI{29.4}{\degree}$, $\alpha_2=\SI{23.9}{\degree}$, $\alpha_3=\SI{19.8}{\degree}$, $\alpha_4=\SI{16.7}{\degree}$, $\alpha_5=\SI{14.4}{\degree}$, which are the same for all sizes of gratings as $\lambda_\mr{max}\le 2\lambda_\mr{min}$ defines a fixed range.
The minimum and maximum deflection angles of the detection bands are plotted exemplarily in a grating with $R_a=\SI{120}{\micro\meter}$ and $d_a=\SI{24}{\micro\meter}$ in Fig.\ref{fig:MultiRC_estimation}b).
Finally, the arclengths of the required output transducers along the Rowland circle (blue line in Fig.\ref{fig:MultiRC_estimation}b) are calculated using $\wideparen{\Delta P_i}=\pi R_x (\alpha_i-\alpha_{i+1})/180$ (or $\wideparen{\Delta P_i}=R_x (\alpha_i-\alpha_{i+1})$ for $\alpha$ in \unit{\radian}) and yield
\begin{itemize}
    \item $R_a=\SI{120}{\micro\meter}$, $d_a=\SI{24}{\micro\meter}$: $\wideparen{\Delta P_1}=\SI{11}{\micro\meter}$, $\wideparen{\Delta P_2}=\SI{8}{\micro\meter}$, $\wideparen{\Delta P_3}=\SI{6}{\micro\meter}$, $\wideparen{\Delta P_4}=\SI{4.4}{\micro\meter}$
    \item $R_b=\SI{60}{\micro\meter}$, $d_b=\SI{12}{\micro\meter}$:  $\wideparen{\Delta P_1}=\SI{5.2}{\micro\meter}$, $\wideparen{\Delta P_2}=\SI{3.8}{\micro\meter}$, $\wideparen{\Delta P_3}=\SI{2.5}{\micro\meter}$, $\wideparen{\Delta P_4}=\SI{2.1}{\micro\meter}$
    \item $R_c=\SI{30}{\micro\meter}$, $d_c=\SI{6}{\micro\meter}$:  $\wideparen{\Delta P_1}=\SI{2.2}{\micro\meter}$, $\wideparen{\Delta P_2}=\SI{1.7}{\micro\meter}$, $\wideparen{\Delta P_3}=\SI{1.2}{\micro\meter}$, $\wideparen{\Delta P_4}=\SI{0.9}{\micro\meter}$.
\end{itemize}
The length of the transducers was reduced by \SI{500}{\nano\meter} to introduce a gap of the same width between the detection regions that were defined seamlessly.
The arc length of the required output transducers starts in the range of those fabricated for the devices demonstrated in the main part of this work (UV-light lithography) and extends to a range where only electron-beam lithography can structure appropriate resist masks.
The transit times $t_t$ of the propagating SW wavefronts (see Appendix~\ref{sec:A_ModelsAndCalcs}) depends on the slope of the dispersion relation (Fig.~\ref{fig:MultiRC_estimation}a left y-axis) and is 
\begin{itemize}
    \item $R_a=\SI{120}{\micro\meter}$, $d_a=\SI{24}{\micro\meter}$: $t_{\mr{t,min}}=\SI{43}{\nano\second}$, $t_{\mr{t,max}}=\SI{54}{\nano\second}$
    \item $R_b=\SI{60}{\micro\meter}$, $d_b=\SI{12}{\micro\meter}$: $t_{\mr{t,min}}=\SI{27}{\nano\second}$, $t_{\mr{t,max}}=\SI{42}{\nano\second}$
    \item $R_c=\SI{30}{\micro\meter}$, $d_c=\SI{6}{\micro\meter}$: $t_{\mr{t,min}}=\SI{21}{\nano\second}$, $t_{\mr{t,max}}=\SI{40}{\nano\second}$.
\end{itemize}

The above estimation shows that the accessible wavelength range for an unambiguous detection is relatively small compared to the full range of wavelengths that could be excited in the grating.
The calculation also shows that the grating periodicity is directly proportional to the wavelength and cannot be freely chosen.
As a result, the remaining degrees of freedom are:
First, the radius $R$ (propagation distance) and arc length of the grating circle compared to the grating period $d$.
Second, the global bias field sets the spectrometer's center frequency (SW wavelength), which can be chosen over a very wide range as long as a sufficiently strong bias field can be provided.
Third, the part of the dispersion that is covered by a single Rowland circle, which is a trade-off between bandwidth, detection resolution, pick-up power at the output transducer, and the ability to fabricate small enough features for gratings with period $d$.

\end{appendices}

\end{document}